\documentclass[aps,pre,twocolumn,groupedaddress,showpacs]{revtex4}
\usepackage{graphicx}
\newcommand{\hrho}{\hat{\rho}}
\newcommand{\ii}{{\rm i}}

\newcommand{\bv}{\mathbf{v}}

\newcommand{\bR}{\mathbf{R}}

\newcommand{\bbr}{\mathbf{r}}

\newcommand{\be}{\beta}
\newcommand{\al}{\alpha}

\newcommand{\sep}{ \ \ \ , \ \ \ }

\newcommand{\beq}{\begin{equation}}
\newcommand{\eeq}{\end{equation}}
\newcommand{\beqn}{\begin{eqnarray}}
\newcommand{\eeqn}{\end{eqnarray}}
\newcommand{\pp}{\partial}
\newcommand{\dd}{{\rm d}}
\newcommand{\ee}{{\rm e}}
\newcommand{\eq}{Eq.\ }
\newcommand{\cf}{c.f.\ }
\newcommand{\ch}{ch.\ }
\newcommand{\eg}{e.g.,\ }
\newcommand{\eqs}{Eqs }
\newcommand{\fig}{Fig.\ }

\newcommand{\cO}{{\cal O}}

\newcommand{\g}{\gamma }
\newcommand{\app}{Appendix\ }
\newcommand{\la}{\langle}
\newcommand{\ra}{\rangle}
\newcommand{\wrt}{{\rm with \ respect \ to \ }}

\begin{document}

\title{Singular perturbation analysis of a reduced model for collective motion:\\ A renormalization group approach
}
\author{Chiu Fan \surname{Lee}
}
\email{cflee@pks.mpg.de}
\affiliation{Max Planck Institute for the Physics of Complex Systems,
N\"{o}thnitzer Str.~38, 01187 Dresden,
Germany
}


\date{\today}

\begin{abstract}
In a system of noisy self-propelled particles with interactions that favor directional alignment, collective motion will appear if the density of particles is beyond a 
critical density. Starting with a reduced model for collective motion, we determine how the critical density depends on the form of the initial perturbation. 
Specifically, we employ a renormalization-group improved perturbative method to analyze the model equations, and show analytically, up to first order in the perturbation 
parameter, how the critical density is modified by the strength of the initial angular perturbation in the system.
\end{abstract}

\maketitle


\section{Introduction}
The interesting phenomena of flocking in animals 
\cite{Toner_AnnPhys05,Couzin_Nature05, Buhl_Science06, Sumpter_PRSB06,Vicsek_a10} and self-organized patterns in motile cells \cite{Tsimring_PRL95,Riedel_Science05,Budrene_Nature91} are currently 
driving the 
intense theoretical study of collective motion among self-propelled particles 
\cite{Vicsek_PRL95,Toner_PRL95, Toner_PRE98,Ramaswamy_EPL03,Gregoire_PRL04,Dossetti_PRE09,Romanczuk_PRL09,Aldana_PRL07,DOrsogna_PRL06,Kruse_PRL04, 
Bertin_PRE06,Peruani_EPJST08,Bertin_JPA09}.  
 Models for collective motion usually involve motile particles that possess alignment interactions and angular noise. Collective motion is then observed  if the density 
of particles increases beyond a certain threshold. We have previously argued that besides density fluctuations, the initial fluctuations in the heading directions of the 
particles constitute another important aspect of the system \cite{Lee_PRE10}.  Here, we  determine the critical density as a function of the initial perturbation 
strength by analyzing a reduced model for collective motion. 
Specifically, we assume that the angular noise strength, $\epsilon$, is small and employ it as the perturbation parameter. We then find that the solution obtained by the
naive perturbation method is plagued by divergences due to the appearances of temporal {\it secular terms} \cite{Holmes_B95,Bender_B99}, which we subsequently eliminate up to order 
$\epsilon$ by the renormalization group method \cite{Chen_PRL94,Chen_PRE96,Matsuba_PRE97, Ei_AnnPhys00, Nozaki_PRE01, Kirkinis_PRE02}.

\begin{figure}\caption{
(a) The form of the initial perturbation with $b=0.1$ and $\eta = 0.5$. (b) The form of the function $R(x)$ (\cf \eq (\ref{fnR})). (c) \& (d)  The temporal evolutions of 
$\alpha$ and $\beta$, with $\epsilon = 0.5$ and $\gamma = 0.9$, respectively.  These results are obtained by performing numerical integrations of \eqs (\ref{main_eq}) with the initial 
square wave perturbation approximated by the function: $b[\tanh (30(x+\eta))-\tanh (30(x-\eta))]$.
}
\label{pic1}
\begin{center}
\includegraphics[scale=.43]{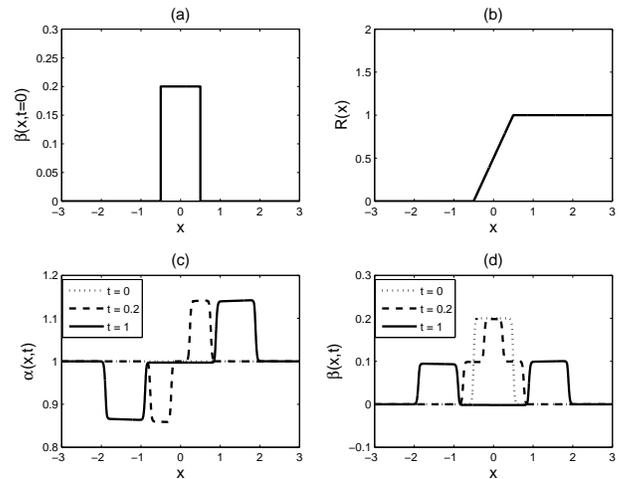}
\end{center}
\end{figure}

\section{Model}

In this work, we consider a minimal model for collective motion in two dimensions introduced in \cite{Peruani_EPJST08}. In this model,  every particle is assumed to have constant 
speed $u$, and that their interactions consist only of a directional alignment mechanism, with interaction strength $\epsilon \g$. Noise, of strength $\epsilon$, is incorporated in the direction of travel. Note that the factor $\epsilon$ in front of the interaction strength is to emphasize that our perturbative treatment will be on both the interaction and noise strengths. Since $\gamma$ is still a free parameter, these two effects can be varied independently. 

Starting from the Fokker-Planck equation describing this model, we have previously argued that at the onset of collective motion, the model equations can be approximated by a finite set of coupled partial differential equations (PDEs) \cite{Lee_PRE10}. 
Here, we will study the simplest set of those coupled PDEs, which corresponds to the following two coupled PDEs:
\beq
\label{main_eq}
\pp_t \al = - 2u \pp_x \be \sep
\pp_t \be = -u\pp_x \al +\epsilon  (\g \al\be -\be)\ .
\eeq
In the above equations, $\al$ corresponds to the local particle density, and $\be$ corresponds to the local vectorial order parameter of the system, i.e., a non-zero $\be$ implies the 
existence of collective motion. Note that the problem is reduced to one dimension as we assume that the initial directional preference of the system is along the $x$-axis.

The derivation of and the approximations involved in the model equations are elaborated in \cite{Lee_PRE10} and we summarize the essential steps in \app \ref{model} for 
completeness.   We are primarily interested in the regime where $\epsilon \ll u$ and $\gamma \sim \cO(1)$. The first condition allows us to study the reduced model 
perturbatively \wrt $\epsilon$, and the second condition is our main interest because it is the region where the transition from disordered motion to collective 
motion occurs, as we will see shortly.

At the mean field level, i.e., if the spatial variations in $\al$ and $\be$ are ignored, the threshold for collective motion is (\cf \eqs (\ref{main_eq})) 
\cite{Peruani_EPJST08,Lee_PRE10}:
\beq
\label{MF}
\rho_c = \g^{-1} \ ,
\eeq
where $\rho_c$ denotes the critical density. 
We have previously argued that such a mean-field picture is incomplete because the initial perturbation to the system should play a major role as well \cite{Lee_PRE10}. 
For instance, let us assume that the initial density is $\rho$, i.e., $\al (x) =\rho$, and $\g <\rho^{-1}$. In other words, no collective motion would be expected 
according to the mean-field description. Let us now consider an initial perturbation in the form of a square wave in  $\be$ with magnitude $2b$. As shown in the next 
section, such a perturbation would induce a density wave of magnitude $\sqrt{2}b$ in $\al$ that travels in the positive $x$ direction (\cf \fig \ref{pic1}(c) and (d)). 
Within the traveling density wave, $\al \sim \rho +\sqrt{2}b$. If
\beq
\label{new_cond}
\rho+\sqrt{2}b > \g^{-1} 
\ ,
\eeq
then the density within the traveling wave is beyond the collective motion threshold. Therefore, according to the mean-field criterion,  we would naively expect that 
such a density wave will be amplified, and thus signals the onset of collective motion. In other words, the critical density may depend on the strength of the initial 
perturbation perturbation. In the next two sections, we will verify this expectation by analyzing the reduced model perturbatively.

\section{Naive perturbative treatment}

We are concerned with the progression of a perturbation to an initially disordered system, and so we are primarily interested in the initial conditions of the form:
\beq
\al(x,t=0) = A \sep
\be (x,t=0) = 2f(x) \ ,
\eeq
where $A$ corresponds to the local density and $f$ is an arbitrary function that describes the initial perturbation in the angular directions of the particles.

We will now consider the model perturbatively by making the following expansions for $\al$ and $\be$:
\beqn
\al&=& \al_0+\epsilon \al_1 +\cO(\epsilon^2)
\\
\be&=& \be_0+\epsilon \be_1+\cO(\epsilon^2) \ .
\eeqn
To the zeroth order, \eqs (\ref{main_eq}) give
\beq
\pp_t \al_0 = - 2u\pp_x \be_0 \sep
\pp_t \be_0 = -u\pp_x \al_0 \ ,
\eeq
and the solutions are the standard traveling wave equations:
\beqn
\nonumber
\al_0(x,t) &=& A+\sqrt{2} \left[ f\left(x-\sqrt{2}ut\right)-f\left(x+\sqrt{2}ut\right)\right]
\\
\label{init0}
\be_0(x,t)&=& f\left(x-\sqrt{2}ut\right)+f\left(x+\sqrt{2}ut\right) \ .
\eeqn
Note that since the units of time and length can be defined arbitrarily, we will use these degrees of freedom to set $u=1/\sqrt{2}$ and $A=1$ from now on.

In the first order in $\epsilon$, \eqs (\ref{main_eq}) lead to
\beqn
\label{a1}
\pp_t \al_1 &=& -\sqrt{2} \pp_x \be_1
\\
\label{b1}
\pp_t \be_1 &=& - \frac{1}{\sqrt{2}}\pp_x \al_1 +  \g \al_0\be_0 - \be_0
\ .
\eeqn
We can eliminate the $\be_1$ term above by combining the partial derivative of \eq (\ref{a1}) \wrt $t$ and the partial derivative of \eq (\ref{b1})  \wrt $x$. We then 
obtain:
\beq
(\pp^2_t - \pp_x^2) \al_1 =\sqrt{2} \pp_x( \be_0- \g \al_0 \be_0 ) \ .
\eeq
The above inhomogeneous wave equation is supplemented by the initial conditions;
\beqn
\al_1(x,t=0) &=& 0 
\\
\label{init1}
\pp_t \al_1 (x,t=0) &=& -\sqrt{2} \pp_x \be_1(x,t=0) =0
\eeqn
where the second equality in \eq (\ref{init1}) follows from \eq (\ref{a1}) and the fact that $\be_1(x,t=0)=0$.
The general solution for $\al_1$ can thus be written as (\eg see \ch 8.2 in \cite{Stakgold_B79}):
\beq
\label{sec_a1}
\al_1 =\frac{1}{\sqrt{2}} \int_0^t \int_{x-(t-t')}^{x+(t-t')} \big[
\pp_{x'}( \be_0-\g \al_0 \be_0 ) \big]\dd x' \dd t' \ .
\eeq
By plugging the expressions for $\al_0$ and $\be_0$ (\cf \eqs (\ref{init0})) into the above integral, we find
\beqn
\label{sing_a1}
\al_1 &=&  \frac{t}{\sqrt{2}} \big[ (\g -1) (f_--f_+)+\g (f_-^2+f_+^2) \big] 
\\
\nonumber
&&+
\frac{\g}{\sqrt{2}} (h_- - h_+) \ ,
\eeqn
where $f_\pm \equiv f(x\pm t)$, $h_\pm \equiv h(x\pm t)$ and the function $h$ is defined by the following ordinary differential equation (ODE):
\beq
\label{hg}
\pp_y h (y)=f^2(y) \ . 
\eeq
Also, we can deduce an expression for $\be_1$ by combining \eq (\ref{sing_a1}) and \eq (\ref{a1}):
\beqn
\label{sing_b1}
\be_1 &=& \frac{t}{2} \big[ (\g -1)(f_-+f_+)+
\g(f_-^2-f_+^2)\big] 
\\
&& +\frac{1-\g }{2}(g_--g_+) \ .
\eeqn
where $g_\pm \equiv g(x\pm t)$ and the function $g$ is defined by the following ODE:
\beq
\label{hg2}
 \pp_y g(y)=f(y) \ . 
\eeq

\subsection*{Example}
Consider the case where the initial perturbation is a square wave of width $2\eta$ and of magnitude $2b$, i.e.,
\beq
\be(x, t=0) = 2b \big[H(x+\eta) -H(x-\eta) \big] \ ,
\eeq
where $H(y)$ is the Heaviside function such that $H(y) =1$ for $y\geq 0$, and zero otherwise.
Note that in order for the perturbative treatment to be valid, the magnitude $b$ has to satisfy the following conditions: $\epsilon b, \epsilon b^2 \ll u=1/\sqrt{2}$ (\cf \eq (\ref{main_eq})).
For this square wave perturbation, we have from \eqs (\ref{hg}) and (\ref{hg2})
\beq
\label{def_hg}
h_\pm = b^2 R(x\pm t)
\sep
g_\pm = b R(x\pm t) \ .
\eeq
where (\cf \fig \ref{pic1}(b))
\beq
\label{fnR}
R(y) = 
\left\{
\begin{array}{ll}
0\ , &  y< -\eta
\\
y\ , &  -\eta \leq y < \eta
\\
2\eta\ , &  y \geq \eta \ .
\end{array}
\right.
\eeq
The temporal evolutions of $\al$ and $\be$ as a result of this initial perturbation are shown in \fig \ref{pic1}(c) and (d).

\section{Renormalization group treatment}

The appearances of $t$ in front of the terms in squared brackets in \eqs (\ref{sing_a1}) and (\ref{sing_b1}) signify that the perturbative solution only makes sense for 
$t\ll \epsilon^{-1}$. In other words, the problem is a singular perturbation problem and the terms 
in squared brackets are called {\it secular terms} \cite{Bender_B99}. To eliminate such  secular terms, we will now follow the renormalization group (RG) method 
introduced in \cite{Chen_PRL94,Chen_PRE96}.

In the RG method, we  first introduce an arbitrary time $\tau$ and split $t$ as $t-\tau +\tau$. 
We then rewrite $f(x\pm t) $ as $B_\pm (\tau) f_\pm$ where $f_\pm \equiv f(x\pm (t-C(\tau))$, such that
\beqn
1 &=& \left[1+\sum_{n=1}^\infty b^\pm_n(\tau) \epsilon^n \right]B_\pm(\tau)
\\
0 &=& C(\tau)+\sum_{n=1}^\infty c_n(\tau) \epsilon^n \ .
\eeqn
The series $\{b_n^\pm \}$ and $\{c_n\}$ are the multiplicative and additive renormalization constants respectively, and they will be chosen order by order in $\epsilon$ 
to eliminate the terms containing $\tau$.

Applying the above expansions to $\al$ and ignoring the non-secular term for the time being, we have for up to order $\epsilon$,
\begin{small}
\beqn
\al
&=&
1+\sqrt{2}\big[(1+b^-_1\epsilon) B_-f_- - (1+b^+_1\epsilon) B_+f_+ \big]
\\
&&
+\epsilon (t-\tau)  \bigg[ \frac{\g  -1}{\sqrt{2}} (B_-f_--B_+f_+)
\\
&&+\g (B_-^2f_-^2+B_+^2f_+^2) \bigg] 
\\
&&
\label{elim}
+\epsilon \tau  \bigg[ \frac{\g -1}{\sqrt{2}} (B_-f_--B_+f_+)+\g (B_-^2f_-^2+B_+^2f_+^2) \bigg] \ .
\eeqn
\end{small}
Our strategy now is to eliminate the term shown in \eq (\ref{elim}) by picking $b^\pm_1$ appropriately. We will again focus on the square wave as our initial 
perturbation, i.e.,
\beq
\label{def_f}
f(y) = H(y+\eta) -H(y-\eta) \ .
\eeq
The virtue of this choice is the  property that $f_\pm^2 = f_\pm$. It is due to this critical property that we can perform the elimination simply by defining the following:
\beq
b_1^\pm = \tau  \bigg[ \frac{\g  -1}{\sqrt{2}} B_\pm +\g B_\pm^2 \bigg]  \ .
\eeq
With $b_1^\pm$ defined as above, $\alpha$ is simplified to
\beqn
\al&=&
1+\sqrt{2}\big[B_-f_- - B_+f_+ \big]
\\
&&+\epsilon (t-\tau)  \bigg[ \frac{\g-1}{\sqrt{2}} (B_-f_--B_+f_+)
\\
&&+\g (B_-^2f_-^2+B_+^2f_+^2) \bigg] +\cO(\epsilon^2) \ .
\eeqn
The same expansion renders $\be$ as follows:
\beqn
\be&=& B_-f_- +B_+f_+
\\
\nonumber
&&+\epsilon
(t-\tau) \bigg[ \frac{\g -1}{2}(B_-f_-+B_+f_+)
\\
&&+
\frac{\g}{\sqrt{2}} (B_-^2f_-^2-B_+^2f_+^2)\bigg]  +\cO(\epsilon^2) \ .
\eeqn
The benefit of the above exercise is that as $\tau$ is arbitrary, the divergence of the original secular terms can be eliminated by picking $\tau$ to be close to $t$. 
Now, the next step is to also eliminate $\tau$ in the above equations. As $\al$ and $\be$ should be independent of how $\tau$ is chosen, their derivatives \wrt $\tau$ 
are zero. Therefore, to first order in $\epsilon$, we arrive at the following differential equations governing $B_\pm$ and $C$:
\beqn
\label{odeB1}
f_-\frac{\pp B_-}{\pp \tau } -f_+\frac{\pp B_+}{\pp \tau } &=& \frac{\epsilon}{\sqrt{2}} \Big[ \g  (B_-^2f_- +B_+^2f_+ ) 
\\
\nonumber
&&+\frac{\g -1}{\sqrt{2}} (B_- f_- -B_+f_+) \Big]
\\
\label{odeB2}
f_-\frac{\pp B_-}{\pp \tau } +f_+\frac{\pp B_+}{\pp \tau } &=& \frac{\epsilon}{\sqrt{2}}  \Big[ \g  (B_-^2f_- -B_+^2f_+ ) 
\\
\nonumber
&&+\frac{\g -1}{\sqrt{2}} (B_- f_- +B_+f_+) \Big] 
\\
\label{odeC}
\frac{\pp C}{\pp \tau }&=& 0
\ .
\eeqn 
\eq (\ref{odeC}) indicates that we can simply set $C(\tau)$ to be zero.  Furthermore,
\eqs (\ref{odeB1}) and (\ref{odeB2})  can be decoupled and we arrive at the following two ODEs:
\beqn
\label{Bpm1}
\frac{\pp B_-}{\pp \tau } &=& \frac{\epsilon}{\sqrt{2}} \left[ \g B_-^2 +\frac{\g  -1}{\sqrt{2}}  B_-  \right]
\\
\label{Bpm2}
\frac{\pp B_+}{\pp \tau } &=& \frac{\epsilon}{\sqrt{2}} \left[ -\g B_+^2 +\frac{\g  -1}{\sqrt{2}}  B_+  \right] \ .
\eeqn
The corresponding solutions are
\beqn
B_+(\tau)&=&\frac{\g -1}{b_+ \ee^{-\epsilon (\g -1) \tau/2} +\sqrt{2} \g  }
\\
B_-(\tau)&=&\frac{\g -1 }{b_- \ee^{-\epsilon (\g -1) \tau/2} -\sqrt{2}\g  } \ ,
\eeqn
where $b_\pm$ are obtained from the initial condition of the problem.

By picking $\tau = t$, we can finally eliminate the original secular terms, and the expressions for $\al$ and $\be$ become 
\beqn
\nonumber
\al&=&1+\sqrt{2}(B_-f_- -B_+f_+)+\epsilon \g (h_--h_+)   +\cO(\epsilon^2)
\\
\nonumber
\be&=& B_-f_- +B_+f_++\epsilon \frac{1-\g}{2} (g_--g_+)+\cO(\epsilon^2)  \ .
\eeqn
Note that in the above solutions, we have reinstalled the non-secular terms $h_\pm$ and $g_\pm$, which are defined in \eqs (\ref{def_hg}). We show that $h_\pm$ and 
$g_\pm$ are not modified under the renormalization procedures in \app \ref{handg}. 

We now consider the temporal evolutions of the system under the designated initial perturbation. 
By assumption, we have at $t=0$,
\beq
\al(x,t=0) = 1
\sep
\be(x,t=0) = 2bf(0) \ .
\eeq
Hence, $B_\pm(0) =b$ and so
\beq
b_+ = \frac{\g -1}{b} - \sqrt{2} \g 
\sep
b_- = \frac{\g -1}{b} + \sqrt{2} \g  \ .
\eeq

Let us focus only on the density wave in $\al$ traveling to the right. The magnitude of the density wave is
\beq
B_-(t) = \frac{b(\g -1)}{((\g -1)+\sqrt{2} b\g ) \ee^{-\epsilon (\g -1)t/2} -\sqrt{2}b \g } \ .
\eeq
If $\g >1$ , then $B_-$ grows with time since the first term in the denominator approaches $\sqrt{2}b \g$ as $t$ grows \cite{footnote}.
In other words, the original perturbation is amplified, which is a signature for the onset of collective motion. This scenario corresponds exactly to our mean-field 
prediction (\cf \eq (\ref{MF})). The more interesting case is that even for $\g  <1$, $B_-$ still grows if
\beq
b > \frac{1-\g}{\sqrt{2} \g } \ .
\eeq
 This is the condition discussed previously and based purely intuition ground (\cf \eq (\ref{new_cond})). In other words, the  critical density in the presence of the 
perturbation  considered is shifted downwards and becomes
\beq
\rho_c = \g^{-1} -\sqrt{2}b  \ .
\eeq

\section{Conclusion}
In this paper, we have demonstrated how the critical density for collective motion can depend on the strength of the initial angular perturbation. Our conclusion is 
based on a renormalization-group improved perturbative treatment of a reduced model for collective motion.  At the simulation level, our results suggest that simulations with different implementations of the initial perturbations may lead to different values of the critical density.

\appendix
\section{Reduced model}
\label{model}
For completeness, we summarize here the essential steps in deriving the equations describing the reduced model studied here \cite{Lee_PRE10}. We consider a minimal model 
for collective motion in two dimensions, where every particle is assumed to have constant 
speed, $u$, and that their interactions consist only of a directional alignment mechanism. Noise, of strength $\epsilon$, is incorporated in the direction of travel. 
Specifically, let there be $N$ particles in a volume of $V$, their equations of motion are:
\beqn
\frac{\dd  \bbr_i}{\dd t} &=& \frac{2u}{\pi} \bv(\theta_i)
\\
\frac{\dd  \theta_i}{\dd t} &=& -\frac{\pp U(\bR, \Theta)}{\pp \theta_i}+\sqrt{2\epsilon}\eta_i(t)
\eeqn
where $1 \leq i \leq N$, $\bR \equiv (\bbr_1, \ldots, \bbr_N)$, $\Theta \equiv (\theta_1, \ldots, \theta_N)$, $\bv(\theta)\equiv (\cos \theta, \sin \theta)$, and the 
noise is assumed to be Gaussian characterized by the following moments:
\beq
\la \eta_i(t) \ra =0 \sep \la \eta_i(t)\eta_j(t') \ra = \delta_{ij} \delta(t-t') \ .
\eeq
Moreover, the alignment interaction is assumed to be of very short range and can thus  be approximated by a delta function:
\beq
U(\bR, \Theta) =- \frac{\epsilon\g}{\pi}\sum_{i<j}\delta^{(2)}(\bbr_i -\bbr_j)  \cos(\theta_i-\theta_j)\ ,
\eeq
where $\epsilon \gamma$ corresponds to the coupling strength.
If we denote the probability distribution of the density of particles in the state $(\bR, \Theta)$ at time $t$ by $f(t, \bR, \Theta)$, then the Fokker-Planck equation 
corresponding  to the system is \cite{Zwanzig_B01}:
\beqn
\nonumber
\frac{\pp f}{\pp t} &=&  \sum_i \bigg\{ \epsilon \frac{\pp^2 }{\pp \theta_i^2}f  -\frac{2u}{\pi}\nabla_{\bbr_i} \cdot [\bv(\theta_i) f] \bigg\}
\\
\label{feq}
&&+\frac{\epsilon \g}{\pi} \sum_{i< j}
\frac{\pp}{\pp \theta_{i}} \left[ \delta^{(2)}(\bbr_i -\bbr_j) 
  \sin(\theta_i-\theta_j)  f \right] \ .
\eeqn
Focusing now on the single-particle density function, $\rho$, where
\beq
\nonumber
\rho(\bbr_1, \theta_1) = \frac{(N!)\int \dd r_2 \cdots \dd r_N \dd \theta_2 \cdots \dd \theta_N f(\bR, \Theta)}{(N-1)!}
\ ,
\eeq
\eq (\ref{feq}) becomes
\begin{small}
\beqn
\nonumber
\frac{\pp \rho(\bbr,\theta)}{\pp t} &=&
\epsilon \frac{\pp^2 \rho(\bbr,\theta)}{\pp \theta^2} 
-\frac{2u}{\pi} \left[
\cos \theta \frac{ \pp \rho(\bbr,\theta)}{\pp x} +\sin \theta \frac{ \pp \rho(\bbr,\theta)}{\pp y} \right]
\\
\label{eq_rho}
&&+\frac{\epsilon \g}{\pi}\frac{\pp }{\pp \theta} \bigg[
 \int \dd \theta' 
\sin (\theta-\theta')
\rho^{(2)}(\bbr,\theta,\bbr,\theta')
\bigg] \ .
\eeqn
\end{small}
where
\beq
\nonumber
\rho^{(2)}(\bbr_1, \theta_1,\bbr_2,\theta_2) =  \frac{(N!)\int \dd r_3 \cdots \dd r_N \dd \theta_3 \cdots \dd \theta_N f(\bR, \Theta)}{(N-2)!}
\ .
\eeq
We now employ the {\it molecular chaos} assumption to close the above hierarchical equation by approximating  $\rho^{(2)}(\bbr, \theta,\bbr,\theta')$ by $\rho(\bbr, 
\theta)\rho(\bbr, \theta')$. Within this approximation, the Fourier transform of \eq (\ref{eq_rho}) with respect to the angular variable, $\theta$, leads to
\beqn
\label{hrho}
\nonumber
\pp_t\hrho_n(\bbr) &=&
-\epsilon n^2 \hrho_n(\bbr)
-u  \Big[ \pp_x \big(\hrho_{n+1}(\bbr)+\hrho_{n-1}(\bbr)\big) 
\\
\nonumber
&&
+\ii \pp_y \big( \hrho_{n-1}(\bbr) - \hrho_{n+1}(\bbr)\big)\Big]
\\
&& -\epsilon \g n\Big[\hrho_{-1}(\bbr) \hrho_{n+1}(\bbr) 
-
\hrho_{1}(\bbr) \hrho_{n-1}(\bbr) \Big] 
\eeqn
where $\rho(\bbr, \theta) = \sum_{n=-\infty}^{\infty} \hrho_n(\bbr)\ee^{-\ii n \theta}$ and $\hrho_n(\bbr)$ is complex.
In \cite{Lee_PRE10}, we have argued that at the onset of collective motion, only the lower modes are important. Therefore, we truncate the above set of infinitely many 
equations by ignoring all $\hrho_n$ such that $n > 1$. Furthermore, if we pick the initial angular perturbation to be directed towards the positive $x$ direction, then 
we need only consider the real part of $\hrho_{\pm 1}$, as the imaginary part of $\hrho_{\pm 1}$ will never be excited \cite{Lee_PRE10}. With these simplifications, the 
original single particle density function is approximated as
\beq
\rho(\bbr, \theta) \simeq \alpha(\bbr) +2\beta(\bbr) \cos \theta \ ,
\eeq
where $\al(\bbr)$ corresponds to the local density, and $\be(\bbr)$ corresponds to the local vectorial order parameter since
\beq
\big\la \bv(\bbr) \big\ra= \big\la (\cos \theta, \sin \theta) \big\ra_{\bbr} = \big(\be(\bbr), 0 \big) \ .
\eeq
From \eq (\ref{hrho}), the reduced model for the onset of collective motion is therefore described by the following equations:
\beq
\pp_t \al = - 2u \pp_x \be
\sep
\pp_t \be = -u\pp_x \al +\epsilon  (\g \al\be -\be) \ .
\eeq

\section{Non-secular terms}
\label{handg}
We consider here the non-secular terms in $\al, \be$ (\cf \eqs (\ref{sing_a1}) and (\ref{sing_b1})). Writing $\al,\be$ in the renormalized forms as follow:
\beqn
\al&=&1+\sqrt{2}(B_-f_- -B_+f_+)+\frac{\epsilon }{\sqrt{2}}h   +\cO(\epsilon^2)
\\
\be&=& B_-f_- +B_+f_++\frac{\epsilon }{2}g+\cO(\epsilon^2)
\eeqn
where $h,g$ are  functions of $x,t$ to be determined by \eqs (\ref{main_eq}). We note that $\pp_t^n B(t) = \cO(\epsilon^n)$, and so to order $\epsilon$, we have from 
\eqs (\ref{main_eq}) the following differential equations governing $h,g$:
\beqn
\label{init_h}
0&=&\sqrt{2}(E_-f_--E_+f_+)+\frac{1}{\sqrt{2}} \pp_t h +\frac{1}{\sqrt{2}}\pp_x g 
\\
0&=& E_-f_--E_+f_++\frac{1}{2}\pp_t g +\frac{1}{2}\pp_x h 
\\
\nonumber
&&- \epsilon \big[(\g -1) (B_-f_- +B_+f_+) +\sqrt{2}(B_-^2f_--B_+^2f_+) \big]
\eeqn
where $E_\pm \equiv \epsilon^{-1} \pp_t B_\pm$. Similar to the derivation of \eq (\ref{sec_a1}), the above equations can be combined to give the following second order 
differential equation for $h$:
\beqn
\big(\pp^2_x - \pp^2_t \big) h &=&4\big( E_-f'_-+E_+f'_+\big)
\\
&&+2\epsilon\Big[ (\g -1) (B_-f_-' +B_+f_+') 
\\
&&+2 (B_-^2f_-' -B_+^2f_+')  \Big]
\\
&=&0 \ ,
\eeqn
where the last identity follows from the definition of $E_\pm$ (\cf \eqs (\ref{Bpm1}) and (\ref{Bpm2})). Given that $h(x,t=0)=0$ and $\pp_t h(x,t=0) =-2(E_-f_--E_+f_+)$ 
(\cf \eq (\ref{init_h})), we have
\beqn
h(x,t) &=& 2 \int_{x-t}^{x+t} [E_+(0)-E_-(0)] f(s) \dd s
\\
&=& -2b^2 \g\int_{x-t}^{x+t} f(s) \dd s
\\
&=& \g (h_--h_+) \ ,
\eeqn
where $f$ is as defined in \eq (\ref{def_f}), and $h_\pm$ are as defined in \eqs (\ref{def_hg}). Moreover, given the expression for $h$, we can now employ \eq 
(\ref{init_h}) to deduce that
\beq
g(x,t) = (1-\g) (g_--g_+)\ ,
\eeq
where $g_\pm$ are also defined in \eqs (\ref{def_hg}).

\acknowledgements
The author thanks Fernando Peruani and Frank J\"{u}licher for helpful discussions.


\begin{thebibliography}{24}
\expandafter\ifx\csname natexlab\endcsname\relax\def\natexlab#1{#1}\fi
\expandafter\ifx\csname bibnamefont\endcsname\relax
  \def\bibnamefont#1{#1}\fi
\expandafter\ifx\csname bibfnamefont\endcsname\relax
  \def\bibfnamefont#1{#1}\fi
\expandafter\ifx\csname citenamefont\endcsname\relax
  \def\citenamefont#1{#1}\fi
\expandafter\ifx\csname url\endcsname\relax
  \def\url#1{\texttt{#1}}\fi
\expandafter\ifx\csname urlprefix\endcsname\relax\def\urlprefix{URL }\fi
\providecommand{\bibinfo}[2]{#2}
\providecommand{\eprint}[2][]{\url{#2}}


  \bibitem{Toner_AnnPhys05}
  J.\ Toner, Y.\ Tu, and S.\ Ramaswamy, Annals of Physics {\bf 318}, 170 (2005).

\bibitem[{\citenamefont{Couzin et~al.}(2005)\citenamefont{Couzin, Krause,
  Franks, and Levin}}]{Couzin_Nature05}
\bibinfo{author}{\bibfnamefont{I.~D.} \bibnamefont{Couzin}},
  \bibinfo{author}{\bibfnamefont{J.}~\bibnamefont{Krause}},
  \bibinfo{author}{\bibfnamefont{N.~R.} \bibnamefont{Franks}},
  \bibnamefont{and} \bibinfo{author}{\bibfnamefont{S.~A.} \bibnamefont{Levin}},
  \bibinfo{journal}{Nature} \textbf{\bibinfo{volume}{433}},
  \bibinfo{pages}{513} (\bibinfo{year}{2005}).
  

\bibitem[{\citenamefont{Buhl et~al.}(2006)\citenamefont{Buhl, Sumpter, Couzin,
  Hale, Despland, Miller, and Simpson}}]{Buhl_Science06}
\bibinfo{author}{\bibfnamefont{J.}~\bibnamefont{Buhl}},
  \bibinfo{author}{\bibfnamefont{D.~J.~T.} \bibnamefont{Sumpter}},
  \bibinfo{author}{\bibfnamefont{I.~D.} \bibnamefont{Couzin}},
  \bibinfo{author}{\bibfnamefont{J.~J.} \bibnamefont{Hale}},
  \bibinfo{author}{\bibfnamefont{E.}~\bibnamefont{Despland}},
  \bibinfo{author}{\bibfnamefont{E.~R.} \bibnamefont{Miller}},
  \bibnamefont{and} \bibinfo{author}{\bibfnamefont{S.~J.}
  \bibnamefont{Simpson}}, \bibinfo{journal}{Science}
  \textbf{\bibinfo{volume}{312}}, \bibinfo{pages}{1402} (\bibinfo{year}{2006}).

\bibitem[{\citenamefont{Sumpter}(2006)}]{Sumpter_PRSB06}
\bibinfo{author}{\bibfnamefont{D.~J.~T.} \bibnamefont{Sumpter}},
  \bibinfo{journal}{Philosophical Transactions of the Royal Society B:
  Biological Sciences} \textbf{\bibinfo{volume}{361}}, \bibinfo{pages}{5}
  (\bibinfo{year}{2006}).

\bibitem{Vicsek_a10}
T.~Vicsek and A.~Zafiris, e-print arXiv:1010.5017.


\bibitem[{\citenamefont{Tsimring et~al.}(1995)\citenamefont{Tsimring, Levine,
  Aranson, Ben-Jacob, Cohen, Shochet, and Reynolds}}]{Tsimring_PRL95}
\bibinfo{author}{\bibfnamefont{L.}~\bibnamefont{Tsimring}},
  \bibinfo{author}{\bibfnamefont{H.}~\bibnamefont{Levine}},
  \bibinfo{author}{\bibfnamefont{I.}~\bibnamefont{Aranson}},
  \bibinfo{author}{\bibfnamefont{E.}~\bibnamefont{Ben-Jacob}},
  \bibinfo{author}{\bibfnamefont{I.}~\bibnamefont{Cohen}},
  \bibinfo{author}{\bibfnamefont{O.}~\bibnamefont{Shochet}}, \bibnamefont{and}
  \bibinfo{author}{\bibfnamefont{W.~N.} \bibnamefont{Reynolds}},
  \bibinfo{journal}{Physical Review Letters} \textbf{\bibinfo{volume}{75}},
  \bibinfo{pages}{1859} (\bibinfo{year}{1995}).

\bibitem[{\citenamefont{Riedel et~al.}(2005)\citenamefont{Riedel, Kruse, and
  Howard}}]{Riedel_Science05}
\bibinfo{author}{\bibfnamefont{I.~H.} \bibnamefont{Riedel}},
  \bibinfo{author}{\bibfnamefont{K.}~\bibnamefont{Kruse}}, \bibnamefont{and}
  \bibinfo{author}{\bibfnamefont{J.}~\bibnamefont{Howard}},
  \bibinfo{journal}{Science} \textbf{\bibinfo{volume}{309}},
  \bibinfo{pages}{300} (\bibinfo{year}{2005}).

\bibitem[{\citenamefont{Budrene and Berg}(1991)}]{Budrene_Nature91}
\bibinfo{author}{\bibfnamefont{E.~O.} \bibnamefont{Budrene}} \bibnamefont{and}
  \bibinfo{author}{\bibfnamefont{H.~C.} \bibnamefont{Berg}},
  \bibinfo{journal}{Nature} \textbf{\bibinfo{volume}{349}},
  \bibinfo{pages}{630} (\bibinfo{year}{1991}).

\bibitem[{\citenamefont{Vicsek et~al.}(1995)\citenamefont{Vicsek, Czir\'{o}k,
  Ben-Jacob, Cohen, and Shochet}}]{Vicsek_PRL95}
\bibinfo{author}{\bibfnamefont{T.}~\bibnamefont{Vicsek}},
  \bibinfo{author}{\bibfnamefont{A.}~\bibnamefont{Czir\'{o}k}},
  \bibinfo{author}{\bibfnamefont{E.}~\bibnamefont{Ben-Jacob}},
  \bibinfo{author}{\bibfnamefont{I.}~\bibnamefont{Cohen}}, \bibnamefont{and}
  \bibinfo{author}{\bibfnamefont{O.}~\bibnamefont{Shochet}},
  \bibinfo{journal}{Physical Review Letters} \textbf{\bibinfo{volume}{75}},
  \bibinfo{pages}{1226} (\bibinfo{year}{1995}).

\bibitem[{\citenamefont{Toner and Tu}(1995)}]{Toner_PRL95}
\bibinfo{author}{\bibfnamefont{J.}~\bibnamefont{Toner}} \bibnamefont{and}
  \bibinfo{author}{\bibfnamefont{Y.}~\bibnamefont{Tu}},
  \bibinfo{journal}{Physical Review Letters} \textbf{\bibinfo{volume}{75}},
  \bibinfo{pages}{4326} (\bibinfo{year}{1995}).

\bibitem[{\citenamefont{Toner and Tu}(1998)}]{Toner_PRE98}
\bibinfo{author}{\bibfnamefont{J.}~\bibnamefont{Toner}} \bibnamefont{and}
  \bibinfo{author}{\bibfnamefont{Y.}~\bibnamefont{Tu}},
  \bibinfo{journal}{Physical Review E} \textbf{\bibinfo{volume}{58}},
  \bibinfo{pages}{4828} (\bibinfo{year}{1998}).

\bibitem{Ramaswamy_EPL03}
S.\ Ramaswamy, R.\ Aditi Simha, and J.\ Toner, EPL {\bf 62}, 196 (2003).


\bibitem[{\citenamefont{Gr\'{e}goire and Chat\'{e}}(2004)}]{Gregoire_PRL04}
\bibinfo{author}{\bibfnamefont{G.}~\bibnamefont{Gr\'{e}goire}}
  \bibnamefont{and}
  \bibinfo{author}{\bibfnamefont{H.}~\bibnamefont{Chat\'{e}}},
  \bibinfo{journal}{Physical Review Letters} \textbf{\bibinfo{volume}{92}},
  \bibinfo{pages}{025702} (\bibinfo{year}{2004}).

\bibitem[{\citenamefont{Dossetti et~al.}(2009)\citenamefont{Dossetti, Sevilla,
  and Kenkre}}]{Dossetti_PRE09}
\bibinfo{author}{\bibfnamefont{V.}~\bibnamefont{Dossetti}},
  \bibinfo{author}{\bibfnamefont{F.~J.} \bibnamefont{Sevilla}},
  \bibnamefont{and} \bibinfo{author}{\bibfnamefont{V.~M.}
  \bibnamefont{Kenkre}}, \bibinfo{journal}{Physical Review E} \textbf{\bibinfo{volume}{79}},
  \bibinfo{pages}{051115} (\bibinfo{year}{2009}).

\bibitem[{\citenamefont{Romanczuk et~al.}(2009)\citenamefont{Romanczuk, Couzin,
  and Schimansky-Geier}}]{Romanczuk_PRL09}
\bibinfo{author}{\bibfnamefont{P.}~\bibnamefont{Romanczuk}},
  \bibinfo{author}{\bibfnamefont{I.~D.} \bibnamefont{Couzin}},
  \bibnamefont{and}
  \bibinfo{author}{\bibfnamefont{L.}~\bibnamefont{Schimansky-Geier}},
  \bibinfo{journal}{Physical Review Letters} \textbf{\bibinfo{volume}{102}},
  \bibinfo{pages}{010602} (\bibinfo{year}{2009}).

\bibitem[{\citenamefont{Aldana et~al.}(2007)\citenamefont{Aldana, Dossetti,
  Huepe, Kenkre, and Larralde}}]{Aldana_PRL07}
\bibinfo{author}{\bibfnamefont{M.}~\bibnamefont{Aldana}},
  \bibinfo{author}{\bibfnamefont{V.}~\bibnamefont{Dossetti}},
  \bibinfo{author}{\bibfnamefont{C.}~\bibnamefont{Huepe}},
  \bibinfo{author}{\bibfnamefont{V.~M.} \bibnamefont{Kenkre}},
  \bibnamefont{and} \bibinfo{author}{\bibfnamefont{H.}~\bibnamefont{Larralde}},
  \bibinfo{journal}{Physical Review Letters} \textbf{\bibinfo{volume}{98}},
  \bibinfo{pages}{095702} (\bibinfo{year}{2007}).

\bibitem[{\citenamefont{D'Orsogna et~al.}(2006)\citenamefont{D'Orsogna, Chuang,
  Bertozzi, and Chayes}}]{DOrsogna_PRL06}
\bibinfo{author}{\bibfnamefont{M.~R.} \bibnamefont{D'Orsogna}},
  \bibinfo{author}{\bibfnamefont{Y.~L.} \bibnamefont{Chuang}},
  \bibinfo{author}{\bibfnamefont{A.~L.} \bibnamefont{Bertozzi}},
  \bibnamefont{and} \bibinfo{author}{\bibfnamefont{L.~S.}
  \bibnamefont{Chayes}}, \bibinfo{journal}{Physical Review Letters}
  \textbf{\bibinfo{volume}{96}}, \bibinfo{pages}{104302}
  (\bibinfo{year}{2006}).

\bibitem[{\citenamefont{Kruse et~al.}(2004)\citenamefont{Kruse, Joanny,
  J\"{u}licher, Prost, and Sekimoto}}]{Kruse_PRL04}
\bibinfo{author}{\bibfnamefont{K.}~\bibnamefont{Kruse}},
  \bibinfo{author}{\bibfnamefont{J.~F.} \bibnamefont{Joanny}},
  \bibinfo{author}{\bibfnamefont{F.}~\bibnamefont{J\"{u}licher}},
  \bibinfo{author}{\bibfnamefont{J.}~\bibnamefont{Prost}}, \bibnamefont{and}
  \bibinfo{author}{\bibfnamefont{K.}~\bibnamefont{Sekimoto}},
  \bibinfo{journal}{Physical Review Letters} \textbf{\bibinfo{volume}{92}},
  \bibinfo{pages}{078101} (\bibinfo{year}{2004}).

\bibitem[{\citenamefont{Bertin et~al.}(2006)\citenamefont{Bertin, Droz, and
  Gr\'{e}goire}}]{Bertin_PRE06}
\bibinfo{author}{\bibfnamefont{E.}~\bibnamefont{Bertin}},
  \bibinfo{author}{\bibfnamefont{M.}~\bibnamefont{Droz}}, \bibnamefont{and}
  \bibinfo{author}{\bibfnamefont{G.}~\bibnamefont{Gr\'{e}goire}},
  \bibinfo{journal}{Physical Review E} \textbf{\bibinfo{volume}{74}}, \bibinfo{pages}{022101}
  (\bibinfo{year}{2006}).

\bibitem[{\citenamefont{Peruani et~al.}(2008)\citenamefont{Peruani, Deutsch,
  and B\"{a}r}}]{Peruani_EPJST08}
\bibinfo{author}{\bibfnamefont{F.}~\bibnamefont{Peruani}},
  \bibinfo{author}{\bibfnamefont{A.}~\bibnamefont{Deutsch}}, \bibnamefont{and}
  \bibinfo{author}{\bibfnamefont{M.}~\bibnamefont{B\"{a}r}},
  \bibinfo{journal}{The European Physical Journal - Special Topics}
  \textbf{\bibinfo{volume}{157}}, \bibinfo{pages}{111} (\bibinfo{year}{2008}).

\bibitem[{\citenamefont{Bertin et~al.}(2009)\citenamefont{Bertin, Droz, and
  Gregoire}}]{Bertin_JPA09}
\bibinfo{author}{\bibfnamefont{E.}~\bibnamefont{Bertin}},
  \bibinfo{author}{\bibfnamefont{M.}~\bibnamefont{Droz}}, \bibnamefont{and}
  \bibinfo{author}{\bibfnamefont{G.}~\bibnamefont{Gregoire}},
  \bibinfo{journal}{Journal of Physics A: Mathematical and Theoretical}
  \textbf{\bibinfo{volume}{42}}, \bibinfo{pages}{445001}
  (\bibinfo{year}{2009}).

\bibitem[{\citenamefont{Lee}(2010)}]{Lee_PRE10}
\bibinfo{author}{\bibfnamefont{C.~F.} \bibnamefont{Lee}},
  \bibinfo{journal}{Physical Review E} \textbf{\bibinfo{volume}{81}},
  \bibinfo{pages}{031125} (\bibinfo{year}{2010}).

\bibitem{Holmes_B95}
M.~H.~Holmes, \emph{Introduction to Perturbation Methods} (Springer, 1995).

\bibitem[{\citenamefont{Bender and Orszag}(1999)}]{Bender_B99}
\bibinfo{author}{\bibfnamefont{C.~M.} \bibnamefont{Bender}} \bibnamefont{and}
  \bibinfo{author}{\bibfnamefont{S.~A.} \bibnamefont{Orszag}},
  \emph{\bibinfo{title}{Advanced Mathematical Methods for Scientists and
  Engineers: Asymptotic Methods and Perturbation Theory}}
  (\bibinfo{publisher}{Springer}, \bibinfo{year}{1999}).

\bibitem[{\citenamefont{Chen et~al.}(1994)\citenamefont{Chen, Goldenfeld, and
  Oono}}]{Chen_PRL94}
\bibinfo{author}{\bibfnamefont{L.~Y.} \bibnamefont{Chen}},
  \bibinfo{author}{\bibfnamefont{N.}~\bibnamefont{Goldenfeld}},
  \bibnamefont{and} \bibinfo{author}{\bibfnamefont{Y.}~\bibnamefont{Oono}},
  \bibinfo{journal}{Physical Review Letters} \textbf{\bibinfo{volume}{73}},
  \bibinfo{pages}{1311} (\bibinfo{year}{1994}).

\bibitem[{\citenamefont{Chen et~al.}(1996)\citenamefont{Chen, Goldenfeld, and
  Oono}}]{Chen_PRE96}
\bibinfo{author}{\bibfnamefont{L.~Y.} \bibnamefont{Chen}},
  \bibinfo{author}{\bibfnamefont{N.}~\bibnamefont{Goldenfeld}},
  \bibnamefont{and} \bibinfo{author}{\bibfnamefont{Y.}~\bibnamefont{Oono}},
  \bibinfo{journal}{Physical Review E} \textbf{\bibinfo{volume}{54}},
  \bibinfo{pages}{376} (\bibinfo{year}{1996}).


\bibitem[{\citenamefont{Matsuba and Nozaki}(1997)}]{Matsuba_PRE97}
\bibinfo{author}{\bibfnamefont{K.~I.} \bibnamefont{Matsuba}} \bibnamefont{and}
  \bibinfo{author}{\bibfnamefont{K.}~\bibnamefont{Nozaki}},
  \bibinfo{journal}{Physical Review E} \textbf{\bibinfo{volume}{56}},
  \bibinfo{pages}{R4926} (\bibinfo{year}{1997}).

\bibitem[{\citenamefont{Ei et~al.}(2000)\citenamefont{Ei, Fujii, and
  Kunihiro}}]{Ei_AnnPhys00}
\bibinfo{author}{\bibfnamefont{S.~I.} \bibnamefont{Ei}},
  \bibinfo{author}{\bibfnamefont{K.}~\bibnamefont{Fujii}}, \bibnamefont{and}
  \bibinfo{author}{\bibfnamefont{T.}~\bibnamefont{Kunihiro}},
  \bibinfo{journal}{Annals of Physics} \textbf{\bibinfo{volume}{280}},
  \bibinfo{pages}{236} (\bibinfo{year}{2000}).

\bibitem[{\citenamefont{Nozaki and Oono}(2001)}]{Nozaki_PRE01}
\bibinfo{author}{\bibfnamefont{K.}~\bibnamefont{Nozaki}} \bibnamefont{and}
  \bibinfo{author}{\bibfnamefont{Y.}~\bibnamefont{Oono}},
  \bibinfo{journal}{Physical Review E} \textbf{\bibinfo{volume}{63}},
  \bibinfo{pages}{046101} (\bibinfo{year}{2001}).

\bibitem[{\citenamefont{Kirkinis}(2008)}]{Kirkinis_PRE02}
\bibinfo{author}{\bibfnamefont{E.}~\bibnamefont{Kirkinis}},
  \bibinfo{journal}{Physical Review E} \textbf{\bibinfo{volume}{78}}, \bibinfo{pages}{032104}
  (\bibinfo{year}{2008}).

\bibitem[{\citenamefont{Stakgold}()}]{Stakgold_B79}
\bibinfo{author}{\bibfnamefont{I.}~\bibnamefont{Stakgold}},
  \emph{\bibinfo{title}{Green's Functions and Boundary Value Problems (Pure and
  Applied Mathematics)}} (\bibinfo{publisher}{John Wiley \& Sons Inc}, 1979).


\bibitem{footnote}
Note that the fact that $B_-$ is allowed to diverge is due to the incompleteness of this reduce model. Specifically, it is due to the truncation of the full set of 
density functions $\{\hrho_n \}$ (\cf \eq (\ref{hrho})). In other words, more higher order modes will have to be incorporated in the consideration as $t$ grows. On the 
other hand, as argued in \cite{Lee_PRE10}, we believe that the consideration of such a reduced model is appropriate at the onset of collective motion, and is beneficial 
for analytical progress.


\bibitem[{\citenamefont{Zwanzig}(2001)}]{Zwanzig_B01}
\bibinfo{author}{\bibfnamefont{R.}~\bibnamefont{Zwanzig}},
  \emph{\bibinfo{title}{Nonequilibrium Statistical Mechanics}}
  (\bibinfo{publisher}{Oxford University Press}, \bibinfo{address}{Oxford},
  \bibinfo{year}{2001}).

\end{thebibliography}
\end{document}